# RF-COMPONENTS EMBEDDED WITH PHOTONIC-BAND-GAP (PBG) AND FISHNET-METAMATERIAL STRUCTURES FOR HIGH FREQUENCY ACCELERATOR APPLICATION

[1]Sara Robak, [1]Daniel Boyden, and [1,2]Young-Min Shin
[1]Department of Physics, Northern Illinois University, Dekalb, IL, 60115, USA
[2]Fermi National Accelerator Laboratory (FNAL), Batavia, IL 60510, USA

*Abstract*

In the development of high efficiency and high gradient RF-accelerators, RF waveguides and cavities have been designed with Photonic Band Gap (PBG) and fishnet-metamaterial structures. The designed structures are comprised of a periodically corrugated channel sandwiched between two photonic crystal slabs with alternating high to low dielectric constants and a multi-cell cavity-resonator designed with fishnet-metamaterial apertures. The structural designs of our interest are intended to only allow an operating-mode or -band within a narrow frequency range to propagate. The simulation analysis shows that trapped non-PBG modes are effectively suppressed down to ~ -14.3 dB/cm, while PBG modes propagated with ~2 dB of insertion loss, corresponding to ~1.14 dB/cm attenuation. The preliminary modeling analysis on the fishnet-embedded cavity shows noticeable improvement of Q-factor and field gradient of the operating mode ($TM_{010}$) compared to those of typical pillbox- or PBG-cavities. Fabrication of the Ka-band PBG-waveguide and S-band fishnet cavity structures is currently underway and they will be tested with a microwave test bench/8510C Network Analyzer and 5.5 MW S-band klystron. These structures can be applied to stable short-bunch formation and mono chromatic radiation in high frequency accelerators.

## INTRODUCTION

Periodic conducting structures have been utilized as active slow wave media interacting with high energy electron beams in a wide range of applications from high gradient linear accelerators to coherent power radiation sources. The distorted dispersion relation of metallic lattices constitutes effective plasmonic modes with negative permittivity (epsilon-negative materials, $\varepsilon < 0$) [1–4]. The non-radiative waves have been widely used in microwave regimes as they can be readily guided in passive components or strongly excited by energy-momentum-coupling with kinetic electrons as an active medium [5]. The confined evanescent waves induce electron bunches when the current density exceeds thermo-dynamic divergence ($\propto f^2$) [6]. This gives rise to a strong coherent electron-photon interaction that is commonly used for high intensity photon generation or high gradient RF acceleration. The spectral regime of the electronic RF devices has rapidly increased from microwave to near terahertz (near-THz) wave regime (0.1–1 THz) as device size can be significantly reduced proportional to the wavelength.

Energy conversion efficiency on the photon-emission or -absorption process is thus noticeably reduced as the optical cycle is decreased [7, 8] Recent trends in the development of THz electron beam devices have gradually transitioned to elliptical sheet beams from round ones as enlarging the beam width either increases radiation intensity or reduces beam current loading. Among various planar slow wave structures (SWSs) recently micro-channels with an asymmetrically aligned double grating array [9] were intensively studied for frequency-tunable coherent radiation source application as they have large instantaneous bandwidth of plasmonic modes that are strongly confined in the beam channel with small ohmic losses. However, this broad band characteristic can lead to unstable oscillation of abnormal modes, which possibly destabilize beam-wave interactions, and perturb normal electronic energy conversion. In particular, the structure becomes heavily overmoded (TE → TEM) as its aspect ratio between x-y transverse dimensions is increased to expand the beam-wave interactive area. It is thus certain that there is a limitation in increasing the aspect ratio of the sheet beam structure due to low-energy mode excitation. A substantial solution to the issue could be obtained from bandgap structures. This paper will present the band-selective planar waveguide combined with the photonic crystal slabs and fishnet-type metamaterial slab. The composite structure permits only non-radiative photonic-band-gap (PBG) modes [2–4] to propagate through the sub-wavelength micro-channel. Comparative numerical analysis and test results on the spectral response and field distribution will be discussed in detail.

## PHOTONIC BAND GAP WAVEGUIDE

In principle, the lowest frequency (cutoff) of a rectangular TE-mode waveguide is determined by waveguide width, so TE modes become closer to TEM modes as the cutoff decreases. Figure 1 shows the $0^{th}$ order dispersion curves ($m = 0$) of TE- and TEM-waveguides. Normalized structural dimensions of the designed models are specified with $a_0/d_0 = 0.75$, $L_0/d_0 = 0.6$, $b_0/d_0 = 0.33$, and $h_0/d_0 = 1.674$, where $d_0$ is the grating period of a unit cell. The surface wall conductivity is defined with oxygen-free high-conductivity (OFHC) copper ($\sigma = 5.8 \times 10^8$ [$\Omega^{-1}$m$^{-1}$]). With an increase in the guide width, the entire eigenfrequencies of the fundamental band (grey solid → grey dashed) are lowered as the other dimensions remain consistent with the ones of the TE mode waveguide. In the $2^{nd}$ TEM waveguide (grey dashed line), the grating height ($L_0$) is shortened to keep the upper cutoff (stop-band) at the same frequency as the TM mode dispersion curve, which further extends the x-y dimensional aspect ratio. The passband of the 2nd TEM SWS (higher aspect ratio: black solid) thus tends to become even more overmoded with larger phase and group velocities. As aforementioned, spatial harmonic components of lower frequency modes are readily coupled with low energy electrons, so that while haunting around in the structure, once excited, they are often self-amplified by phase-mismatching condition, which can cause strong perturbation to electron-photon coupling.

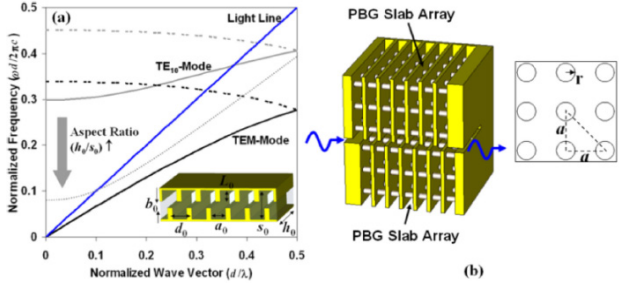

Figure 1: (a) Dispersion graph of the overmoded planar SWS. Inset is the schematic drawing of the staggered double grating arrayed waveguide with the dimensional parameters (TEM-mode) (b) 3D drawing of the PBG-slab-arrayed waveguide and unit cell of the designed PBG slab.

Embodiment of PBG elements in guided wave structures significantly reduces background noise level, including non-PBG modes, in the dynamic spectral range. In Fig. 1(b), the top and bottom gratings are replaced with the arrayed PBG slabs, which consist of alumina ($Al_2O_3$: $\varepsilon_r = 9.4$, $\tan\delta = 0.0004$) rods. In the crystal designed with the lattice constant of $a/\lambda = 0.3$ and filling ratio of $r/a = 0.25$, normalized global band-gap emerges at f (= $\omega d_0/2\pi c$) = 0.25 –0.3. As depicted in Fig. 1(b), like an ordinary confined slow wave structure, electromagnetic (EM) waves bouncing back from the crystal lattices in the band-gap spectra that satisfy the Bragg condition propagate through the channel between the two PBG slab-arrays with strong field confinement. On the other hand, non-resonant eigenmodes outside the band gaps are freely radiated through the lattices, which are thereby strongly suppressed in the channel. Figure 2 shows dispersion curves, (a), of the TEM waveguide with the staggered vane arrays and transmission spectra of the photonic crystal slab, (b), and the PBG-slab embedded TEM waveguide, (c), with respect to two filing ratios, $r/a = 0.2$ and 0.3. In Fig. 3(b), the crystal slab arrays with $r/a = 0.2$ and 0.3 have three band gaps below the coalesced mode (upper cutoff) of the fundamental passband (a). Therefore, only the eigenmodes in the band gaps are allowed to propagate through the beam channel. Figure 3(c) shows an apparent emergence of three spectral bands with zero insertion loss which are exactly matched with the band gaps in Fig. 3(b). The background loss level of radiative non-PBG modes is about _45 dB that is low so as not to cause an abnormal beam-wave interaction. In Figs. 2(b) and 2(c), it is readily noticeable that the band-gap widths of $r/a = 0.3$ is only about half of widths of $r/a = 0.2$, which implies that increase of the rod size relative to the lattice constant leads to raising population density of guided photonic eigenstates with narrowing their intrinsic bandwidths. This indicates that the active spectral range and dispersion characteristics of confined waves are primarily determined by a lattice constant and a filling ratio of the photonic crystal filter rather than by the grating dimensions.

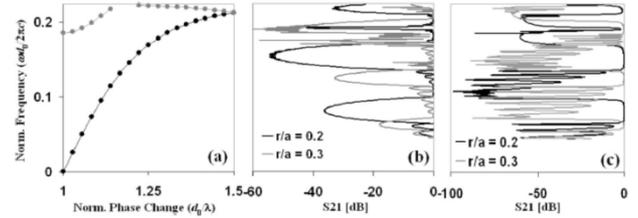

Figure 2: (a) $0_{th}$ order dispersion graph of TEM-mode SWS and transmission spectra of the PBG-slabs (b), and the PBG slab-arrayed SWS (c), of $r/a = 0.2$ (black) and 0.3 (grey).

The simulation models in Figures 2 and 3 are designed to be the ideal TEM-mode waveguide with the magnetic boundaries in the major transverse axis that appear impractical in actual application. The actual device model is thus designed with the oversized TE-mode waveguide that has an electric boundary. Figure 3(a) is the dispersion curves of the two slow waveguide structures embedded with the grating arrays (grey) and the photonic crystal slab arrays (red) that are procured by the finite-integral-technique (FIT) eigenmode solver.

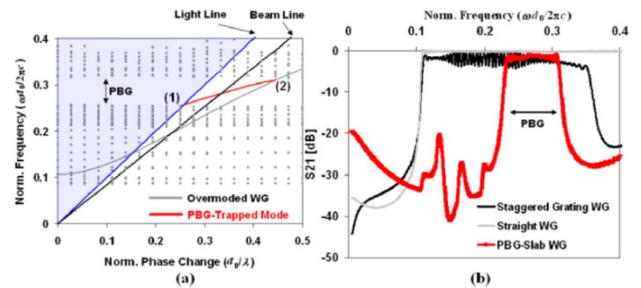

Figure 3: (a) $0_{th}$ order dispersion graphs (band-diagram) of the PBG-slab-arrayed SWS and TEM-mode SWS. (b) Transmission spectra (S21) of staggered grating waveguide, straight waveguide, and PBG-slab-arrayed waveguide.

The blue-shaded and white colored areas represent the photonic zone (far-field, $k < 2\pi/\lambda$) and plasmonic one (near-field, $k > 2\pi/\lambda$), respectively. The plotted points indicate all scattering modes corresponding to a phase change per period of the staggered corrugation. These unguided waves travel around and radiate in free space, and therefore have no interaction with the electron beam. In Fig. 3(b), the global band-gap noticeably emerges from f = 0.265 to f = 0.31 at either the photonic or plasmonic wave area. There is no other photon mode than interactive plasmon modes in the first forbidden band. The channel confined interaction modes disappear beyond the stop-band as they radiate to annihilate owing to the non-resonant matching condition with the crystal lattices. Note that the PBG-modes are considerably more dispersive than fundamental eigenmodes of the oversized waveguide (grey line). Figure 4 (a) shows a K-band test device machined and assembled with two PBG slabs. The reflection and transmission coefficients ($S_{11}$ and $S_{21}$) are measured by a vector network analyzer (HP8510C), as shown in Fig. 4(b). The experimental data, well agreeing with simulation result, clearly show the PBG band in the transmission spectrum. The transmission spectrum of the PBG band is ~ 25 dB higher than that of the non-resonant band.

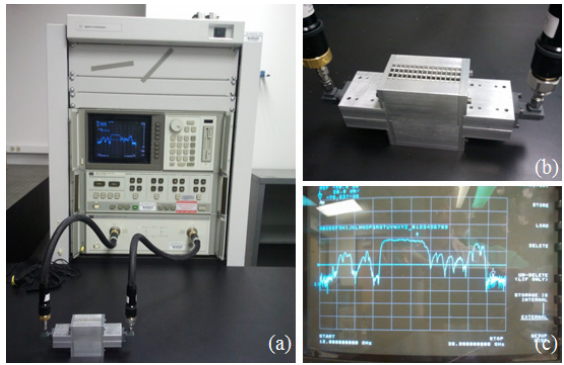

Figure 4: (a) Experiment setup with HP8510C vector network analyzer (b) K-band test device (c) transmission spectrum ($S_{21}$) of 12 – 38 GHz.

## FISHNET-METAMATERIAL RF CAVITY

A higher frequency SRF cryomodule will make a great benefit that can save construction and operation costs, in addition to reduce a physical size of facility and to increase accelerating gradients. Also, operating at high frequency and low bunch charge reduces the risks of brightness degradation during beam transport. However, presence of higher-order-mode (HOMs) wakefields in the high-Q cavities possibly decreases RF energy conversion efficiency, plus inducing thermal heating. Especially, for high-duty-factor machines the heat is not easily extracted, which possible causes deformation or even quenching. The wakefields, normally proportional to a cube of the frequency, can significantly reduce luminosity and cause a beam breakup (BBU), which can severely destabilize acceleration and collision of beams in the long-pulse or CW mode operation. A fishnet structure [ref], consisting of the sandwiched metal-dielectric-metal layer [2 – 4], is one of the simplest double negative meta-materials with negative refractive indexes at resonances. The negative field response can efficiently remove HOMs from a beam interaction area over the broad spectral range. We designed two types of pillbox cavities with fishnet absorbers. Figure 5 shows the cavities strongly suppress a level of HOM wakefields down to - 30 ~ 40 dB in the transmission spectrum. Note that the accelerating mode becomes much narrower with the fishnet-embedded cavities. It is found that a small size cavity with the same physical dimensions with an accelerating mode ($TM_{010}$) cavity can be used to suppress or filter HOMs with the structure (over-sizing cavity volume is unnecessary). We also observed that implementing the structure in the cavity doubles Q-values (loaded Q, $Q_L$). It is expected that negative dispersion response of the novel RF component have potential to advance SRF technology and we will look into its field characteristics and thermal properties of HOMs by simulation and experimentation.

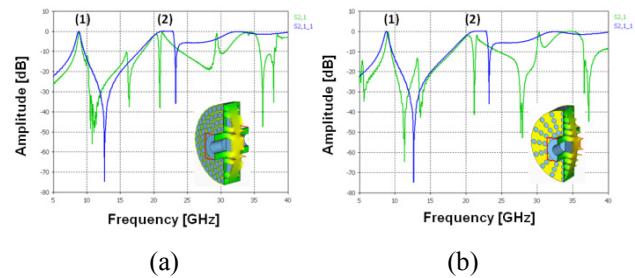

(a)           (b)

Figure 5: Transmission spectrums of fishnet metamaterial RF cavities (a) type-1 and (b) type-2 ((1) TM010 and (2) HOM)

## CONCLUSION

The staggered double PBG-slab arrays enable over-moded planar slow waveguides to induce monochromatic wave propagation of a single plasmonic mode within a narrow band-gap. This highly selective band filtering noticeably improves spectral purity of plasma-interactive evanescent waves in the over-moded slow waveguides, free from the instability problems. A fishnet meta-material absorber strongly suppresses HOM wakefields down to – 30 ~ 40 dB range in a high-Q RF cavity. The quasi-optical mode-filtering/damping/absorbing devices support efficient energy conversion, which can be utilized as a passive or active component for various optoelectronic and electron beam devices.